\documentclass[10pt,a4]{article}
\usepackage{graphicx}
\usepackage{epsfig}
\usepackage{fullpage}
\usepackage{mathrsfs}
\newcommand{\eqref}[1]{(\ref{#1})}

\begin{document}
\begin{center}
{\bf QHJ route to multi-indexed exceptional Laguerre polynomials and corresponding rational potentials}\\
S. Sree Ranjani \footnote{Email: s.sreeranjani@gmail.com}\\
Faculty of Science of Technology, ICFAI Foundation for Higher Education \\ (Declared as Deemed-to-be University u/s 3 of the UGC Act 1956)\\ Dontanapally, Hyderabad, India, 501203.\\
\end{center}

\begin{center}
{\bf Abstract}
\end{center}
A method to construct multi-indexed exceptional Laguerre polynomials using isospectral deformation technique and quantum Hamilton-Jacobi (QHJ) formalism is presented. We construct generalized superpotentials using singularity structure analysis which lead to rational potentials with multi-indexed polynomials as solutions. We explicitly construct such rational extensions of the radial oscillator and their solutions, which involve exceptional Laguerre orthogonal polynomials having two indices. The exact expressions for the $L1$, $L2$ and $L3$ type polynomials, along with their weight functions are presented. We also discuss the possibility of constructing more rational potentials with interesting solutions. \\

\noindent
{\bf keywords} Exceptional orthogonal polynomials, multi-indexed exceptional polynomials, exactly solvable models, rational potentials, shape invariance, isospectral deformation,  quantum Hamilton-Jacobi formalism, quantum momentum function.\\

\noindent
{\bf 1. Introduction}\\

The previous decade saw a lot of interesting work in the areas related to Sturm-Loiville's theory and orthogonal polynomials owing to the discovery of exceptional orthogonal polynomials (EOPs) \cite{kam1}. The subsequent construction of rational potentials with these polynomials as solutions \cite{quesne},  has led to a renewed interest in exactly solvable quantum mechanics. Various studies delved into the different aspects of these EOP systems like their classification \cite{angeles}, spectral analysis \cite{kamL3}, the structure of the zeros and other interesting properties \cite{sasaki_sigma} - \cite{kuij}. Simultaneously  different methods were developed to construct these polynomials and the rational potentials \cite{kam3} -\cite{sreeSIP}. Over the years these new polynomials appeared in connection with nonlinear oscillators \cite{cari}-\cite{fellows} and superintegrability \cite{qu_super_integ}. They can also be found in the context of quantum information theory \cite{Qinfo}, discrete quantum mechanics \cite{dis_qm} and Schr\"odinger equation with position dependent mass and other studies \cite{midya}-\cite{sigma}. 

Another significant development has been the construction of multi-indexed EOPs using methods like Krein-Adler transformations, multi-step Darboux transformation and higher order SUSY \cite{multi} - \cite{{multi-sigma}}. The explicit construction of two indexed Laguerre polynomials has been presented in \cite{multi1}. The construction of these  generalized families of multi-indexed polynomials is more complicated owing to the weight regularity problem \cite{disconj}. These polynomials have a complex structure and are currently being studied \cite{multi}, \cite{multi-sigma}. Compared to these the single indexed EOPs are well studied.  Here, we have generalized families of the exceptional Hermite, three families of exceptional Laguerre, namely $L1, L2, L3$  and two families of exceptional Jacobi polynomials, $J1$ and $J2$ \cite{sasaki_sigma} - \cite{david}. These polynomials are characterized by  the codimension index $m$, which gives the number of gaps in the sequence of polynomials and can take values $1,2,3,\dots$. Therefore $X_m$ EOP sequence implies a polynomial sequence with $m$ number of gaps and different $m$ gives a different order of exceptional polynomials.  For the review of the EOPs and exactly solvable quantum mechanics we refer to \cite{sasaki-univ}.

 The existence of the single and multi-indexed EOPs has extended the class of orthogonal polynomials and widened the scope of the Bochner's theorem \cite{angeles}, \cite{boch}  on classical orthogonal polynomials (COPs) \cite{osci}, \cite{erd}. The differential equations of the EOPs have rational coefficients and the polynomials form a complete set with respect to a rational weight function. For a well defined spectral problem, these weight functions should be well behaved and should not have any singularities in the domain of orthonormality. Therefore we need to ensure that the weight regularity problem does not arise while constructing these polynomials \cite{kamL3}. 

In a recent paper \cite{sreeSIP}, we used the technique of isospectral deformation \cite{dutta}, \cite{khare_book} to construct the rational potentials having $X_m$ EOPs as solutions. In the present study, we couple this method with the simple, yet powerful techniques of the quantum Hamilton-Jacobi formalism (QHJF), to construct rational potentials having multi-indexed EOPs in their solutions.  We make use of the fact that for the zero energy state, the superpotential is equivalent to the quantum momentum function (QMF) of the given potential. Within the framework of the QHJF, the singularity structure of the ES potentials and of the corresponding QMFs are well understood. This understanding is exploited here to obtain the required results. 

For a given ES potential, using isospectral deformation, we construct a generalized superpotential such that one of the partners is the original potential and the other is its rational extension. The complete form of the new superpotential is obtained by doing the singularity structure analysis. In the process we also obtain all the possible generalized superpotentials associated with the original potential.  We use these to construct the rational extensions of the potential and analyze their solutions. In this paper, we consider the rational oscillator potential and explicitly construct the rational potentials with solutions in terms of the two indexed exceptional Laguerre polynomials of $L1$, $L2$ and $L3$ type. We also discuss the construction of a hierarchy of rational potentials with multi-indexed EOPs as solutions, by iteratively applying this method.

The presentation of the paper is as follows. In section $2$, we give a brief introduction to the supersymmetric quantum mechanics (SUSYQM), followed by a description of isospectral deformation technique for the first and the second iterations in section $3$. In section $4$,  we summarize the  QHJF and its connection with the SUSYQM. In section $5$, we discuss the construction of the rational potentials. In section $5.1$, we summarize our results on the three generalized families of rational potentials associated with the radial oscillator and their solutions involving $X_m$ exceptional Laguerre polynomials. In section $5.2$, we perform the second iteration explicitly and construct rational potentials and their solutions involving  two indexed EOPs. This is followed by a discussion of the results in section $6$ and concluding remarks in section $7$. For convenience, we have put $(\hbar=2m=1)$ through out this paper.\\

\noindent
{\bf 2 Supersymmetric quantum mechanics }					\\
       In SUSYQM \cite{khare_book}, we have a pair of supersymmetric  partner potentials
\begin{equation}
V^{\pm}(x)= W^2(x) \pm \partial_x W(x),  \label{partners}
\end{equation}					
where 
			\begin{equation}
W(x)=-\frac{d}{dx}\ln \psi_0^-(x), \label{suppot}
\end{equation}
is the the superpotential associated with the pair. Here $\psi_0^-(x)$ is the ground state wave function of $V^-(x)$. If  $\psi_0^-(x)$ is normalizable with $E_0^-=0$ and $A\psi^-_0(x)=0$, then SUSY is known to be exact between the corresponding partners. In this case, $\psi_0^+(x)$ is nonnormalizable and the partners are isospectral except for the ground states. In this case the wave functions, $\psi^{\pm}_n(x)$, of the partners are related by
\begin{equation}
\psi_{n+1}^-(x)=A^{\dag}\psi_n^+(x)  \,\,\,;\,\,\, \psi_n^+(x)=A \psi_{n+1}^-(x),  \label{wfsexact}
\end{equation}
where $n=0,1,2,\dots$ and
\begin{equation}
A=\frac{d}{dx}+W(x)\,\,\,;\,\,\,A^{\dag}=-\frac{d}{dx}+W(x) \label{intertwine}
\end{equation}
are the intertwinning operators. For all the ES models the wave functions are of the form
\begin{equation}
\psi_n^-(x)= \psi_0^-(x) P_n(x),               \label{wf1}
\end{equation}
where $P_n(x)$ is a COP. Using \eqref{suppot}, we obtain
\begin{equation}
\psi_n^-(x)= \exp \left(-\int W(x)dx \right) P_n(x).               \label{wf2}
\end{equation}
On the contrary, SUSY is said to be broken, if $W(x)$ leads to nonnormalizable ground states for both the partners. Here, both the partners are isospectral including the ground states. The relation between the eigenfunctions of the partners is 
\begin{equation}
\psi_{n}^-(x)=A^{\dag}\psi_n^+(x)  \,\,\,;\,\,\, \psi_n^+(x)=A \psi_{n}^-(x).  \label{wfsbroken}
\end{equation}
Using the above equations, we can construct an hierarchy of ES potentials and their solutions from $V^-(x)$ and its solutions.  The existence of the rational potentials has led to the idea of SUSY being nontrivial \cite{chait}, which will be elaborated in the later sections. \\

\noindent
{\bf Shape Invariance}\\
The partners  $V^{\pm}(x)$ are known to be translationally shape invariant potentials (SIPs), if					
\begin{equation}
V^+(x,a_0)= V^-(x, a_1) + R(a_0),\label{SI}
\end{equation}
where $a_0$ is the potential parameter, with $a_1$ and $R(a_0)$ being functions of $a_0$. In these cases, solutions of  $V^+(x,a_0)$ can be obtained in a simple way using
\begin{equation}
\psi^+_n(x)=\psi^-_n(x,a_0 \rightarrow a_1),    \label{SIsols}
\end{equation}
without having to use the intertwinning operators. In one dimension, shape invariance is the sufficient condition for the potentials to be exactly solvable \cite{ged}. Therefore all the one dimensional ES potentials are shape invariant (SI). In fact this condition can be used to construct all the known ES potentials, including the new rational potentials \cite{sip}.  For more details on SUSYQM, we refer the reader to \cite{khare_book} and the references therein.
 
It is well known that a given potential can have more than one superpotential associated with it. Each superpotential, $W_i(x)$, leads to partner potentials $V^{\pm}_i(x)$, where every $V^-_i(x)= V^-(x) + C_i$, with $C_i$ being a constants. SUSY  may be exact or broken between the partners depending on whether $W_i(x)$ leads to normalizable ground state or not. Moreover the solutions of all the partners, $V_i^+(x)$, associated with $V_i^-(x)$,  can be obtained from \eqref{SIsols}, where $\psi_n^-(x)$ is substituted from \eqref{wf2} with suitable values of $a_1$ in each case. In the discussion of the rational potentials and the EOPs, the various superpotentials associated with $V^-(x)$, the corresponding partners and their solutions play a crucial role in the isospectral deformation method.  \\ 

\noindent
{\bf 3 Isospectral Deformation}\\
     Given the supersymmetric partners $V^{\pm}_i(x)$ and the corresponding superpotential $W_i(x)$,  we construct a general superpotential 
\begin{equation}
\tilde{W}_i(x)=W_i(x) +\phi_1(x), \label{wtil}
\end{equation}
by demanding
\begin{equation}
\tilde{V}^+_i(x) = V^+_i(x)+R_1,  \label{isoshift}
\end{equation}
where $R_1$ is a constant to be determined. The partner potentials associated with $\tilde{W}(x)$ are 
\begin{equation}
\tilde{V}^{\pm}_i(x)=\tilde{W}^2_i(x) \pm\partial_x\tilde{W}_i(x).  \label{vtilpm}
\end{equation}
We can determine $\phi_1(x)$ by writing \eqref{isoshift} in terms of the two superpotentials,
\begin{equation}
\tilde{W}^2_i(x)+\partial_x\tilde{W}_i(x)=W^2_i(x) + \partial_x W_i(x)+R_1.  \label{}
\end{equation}					
Making use of \eqref{wtil} gives
 \begin{equation}
\phi^2_1(x)+2 W_i(x) \phi_1(x)+ \partial_x\phi_1(x)-R_1=0,    \label{phieqn}
\end{equation}					
which is nothing but a Riccati equation. Linearizing it using the Kole-Hopf transformation,
\begin{equation}
\phi_1(x)=\frac{\partial_xP(x)}{P(x)}, \label{phi}
\end{equation}			
gives	the second order differential equation
\begin{equation}
\partial_x^2P(x)+2W_i(x)\partial_x P(x)-R_1 P(x)=0.  \label{Peqn}
\end{equation}
For all the ES models studied, by demanding that $P(x)$ be an $m^{th}$ degree polynomial, with $m=0,1,2\dots$, a suitable point canonical transformation (PCT) reduced \eqref{Peqn} to a second order differential equation of one of the COP, with a condition on $R_1$. Thus $P(x)$ coincides with a COP denoted by $P_m^{\alpha_i}(r)$, where the index $i$ in $\alpha_i$ gives the correspondence to the index $i$ in $W_i(r)$.  Thus we obtain 
\begin{equation}
\tilde{W}_i(x)=W_i(x) +\frac{\partial_xP_m^{\alpha_i }(x)}{P^{\alpha_i}_m(x)}. \label{wtild}
\end{equation}
By substituting different $W_i(x)$ associated with $V^-(x)$ in \eqref{Peqn}, we obtain different $P^{\alpha_i}_m(x)$ leading to different $\tilde{W}_i(x)$.   \\

\noindent
{\bf Construction of rational potentials and their solutions}\\
The partners $\tilde{V}^{\pm}_i(x)$ associated with  $\tilde{W}_i(x)$  can be constructed using
\begin{equation}
\tilde{V}^{\pm}_i(x)=\tilde{W}_i(x)\pm \partial_x\tilde{W}_i(x).
\end{equation}
By definition \eqref{isoshift}, $\tilde{V}^+_i(x)$ is same as $V^+_i(x)$, but shifted by $R_1$, but its partner $\tilde{V}^-_i(x)$ turns out to be a distinct new potential of the form
\begin{equation}
\tilde{V}^-_i(x) =  V^-_i(x)-2 \partial_x\phi_1(x)+R_1.  \label{vtil}
\end{equation}
This in terms of $P^{\alpha_i}_m(x)$ becomes
\begin{equation}
\tilde{V}^-_i(x) = V^-_i(x)-2 \partial_x\left (\frac{\partial_x P^{\alpha_i}_m(x)}{P_m^{\alpha_i}(x)}\right) +R_1.  \label{vtilminus}
\end{equation}
From the above equations, it is clear that for each value of $m$, we get a different family of potentials and these $\tilde{V}^-_i(x)$ are all rational extensions of the original potentials  $V^-_i(x)$.  We call these the first generation rational potentials. In the limit $m \rightarrow 0$, $\tilde{W}_i(x)\rightarrow W_i(x)$  and $\tilde{V}^-_i(x) \rightarrow V^-_i(x)$, since for all COP sequences $P^{\alpha_i}_0(x)=1$. The supersymmetry machinery available, allows us to construct the  eigenvalues and eigenfunctions for $\tilde{V}^-_i(x)$, without having to solve the Schr\"odinger equation. The fact that $\tilde{V}^+(x) \equiv V^+(x)$ implies their solutions are also equivalent. Therefore, the $n^{th}$ excited state of $\tilde{V}^+(r)$, 
\begin{equation}
 \tilde{\psi}^+_n(x) = \psi^+_n(x),
\end{equation}
apart from a normalization constant. We make use of the intertwinning operators
\begin{equation}
\tilde{A}= \frac{d}{dx}+\tilde{W}(x)\,\,\,;\,\,\, \tilde{A}^{\dag}= -\frac{d}{dx}+\tilde{W}(x), \label{tildeops}
\end{equation}
defined for $\tilde{W}(r)$ and $\tilde{V}^{\pm}(r)$ to construct $\tilde{\psi}_{n}^-(x)$. Using \eqref{wfsbroken}, we have
\begin{equation}
\tilde{\psi}_{n}^-(x)=\tilde{A}^{\dag}\tilde{\psi}^+_{n}(x)\,\,;\,\, n=1,2\dots \label{wf}
\end{equation}
For all the rational extensions constructed, the wave functions are of the forms
\begin{equation}
\tilde{\psi}_{n}^-(x)= \frac{\psi_0^-(x)}{P^{\alpha_i}_m(x)}  \mathcal{P}^{\alpha_i}_{m,n}(x),   \label{wf3}
\end{equation}
where $\mathcal{P}^{\alpha_i}_{m,n}(x)$ are the $X_m$ EOPs of codimension $m$, the degree of $P^{\alpha_i}_m(x)$. Here $\psi_0^-(x)$ is the normalizable ground state of $V^-(x)$.  From the above equation it can be clearly seen that for $\tilde{\psi}_{n}^-(x)$ to be an acceptable wave function, $P_m^{\alpha_i}(x)$ should not have any zeros in the domain of orthonormality. The choice of suitable $W_i(x)$ in \eqref{Peqn} takes care of this.

The eigenvalues of $\tilde{V}^-(x)$  can also be obtained in a simple way. From \eqref{isoshift}, it is obvious that $\tilde{E}^+_n= E^+_n+R_1$ and since $\tilde{V}^{\pm}(x)$ are strictly isospectral, the eigenvalues of $\tilde{V}^-(x)$ will be
\begin{equation}
\tilde{E}^-_n= E^+_n+R_1. \label{energies}
\end{equation}
In \cite{sreeSIP}, we constructed the different generalized rational potentials and their solutions for the radial oscillator and trigonometric P\"oshl-Teller potentials using this method. This can be used to rationally extend all known shape invariant potentials in one dimension and show that their solutions are in terms of the single indexed EOPs. 
In the next section, we perform a second iteration of isopectral deformation and construct rational extensions of $\tilde{V}_i^-(r)$ and construct their solutions.\\

\noindent
{\bf 3.2 Second iteration of isospectral deformation}\\
  We begin with  $\tilde{W}_i(x)$ and construct a new generalized superpotential 
\begin{equation}
\bar{W}_i(x)=\tilde{W}_i(x) + \phi_2(x),   \label{wbar}
\end{equation}
by demanding that 
\begin{equation}
\bar{V}^-(x)=\tilde{V}^-(x) +R_2.  \label{vbar}  
\end{equation}
Here the constant $R_2$ and and $\phi_2(x)$ need to be determined. Proceeding as in the first iteration, the equation for $\phi_2(x)$ turns out to be
\begin{equation}
\phi^2_2(x)+2 \tilde{W}_i(x) \phi_2(x) - \partial _r\phi_2(x)-R_2=0,    \label{phi2eqnor}
\end{equation}
which again is a Riccati equation for each $\tilde{W}_i(x)$. The simple Kole-Hopf transformation, used in the first iteration, is not sufficient to give us the complete structure of $\phi_2(x)$. We show that the the QHJ formalism provides the necessary inputs to fix $\phi_2(x)$. Once $\phi_2(x)$ is determined, we can construct $\bar{W}_i(x)$, which in turn can be used to construct the partners $\bar{V}^{\pm}_i(x)$. Again by definition \eqref{vbar}, $\bar{V}^-_i(x)$  are same as $\tilde{V}^-_i(x)$, but $\bar{V}^+_i(x)$ will be distinct potentials of the form
\begin{equation}
\bar{V}^+_i(x) =  \tilde{V}^+_i(x)+2\partial_x\phi_2(x) +R_2,         
\end{equation}
which is a rational extension of $\tilde{V}^+_i(x)$. Using \eqref{isoshift}, we can write
 \begin{equation}
\bar{V}^+(x) = V^+_i(x)+2\partial_x\phi_2(x) +R_2 +R_1,         \label{vplusbar}
\end{equation}
which shows that it is a rational extension of the $V_i^+(r)$, obtained after a second iteration of isospectral deformation. We show that for these extensions, the rational terms are a combination of  the logarithmic derivatives of the COPs and $X_m$ EOPs. We call these the second generation rational potentials.  As in the first iteration, we can make use of the fact that $\bar{\psi}^-_n(x) = \tilde{\psi}^-_n(x)$ apart from a normalization constant and construct the solutions of the $\bar{V}^+_i(x)$ using
\begin{equation}
\bar{\psi}^+_n(x)= \bar{A} \bar{\psi}^-_n(x),    \label{psibar}
\end{equation}
where
\begin{equation}
\bar{A} =\frac{d}{dx} + \bar{W}(x)\,\,\,;\,\,\, \bar{A}^{\dag}=-\frac{d}{dx} + \bar{W}(x)   \label{abar}
\end{equation}
are the intertwinning operators connecting the solutions of $\bar{V}^{\pm}(x)$. Substituting \eqref{wf3} in \eqref{psibar}, we obtain eigenfunctions of the form  
\begin{equation}
\bar{\psi}^+_n(x)= \frac{\psi_0^-(x)}{\mathcal{P}^{\alpha_i}_{m,n^{\prime}}(x)} \mathcal{Q}_N(x), \label{wf4}
\end{equation}
and we show that $ \mathcal{Q}_N(x)$ is an EOP with two indices. It should be noted that for the eigenfunctions to be well behaved, the polynomial appearing in the denominator should not have any zeros in the orthonormality interval. In the next section, we show that the singularity structure analysis of $\bar{W}(x)$ ensures such behaviour. 
 \noindent
The eigenvalues in this case will be
\begin{equation}
\bar{E}^+_n=\tilde{E}^-_n+R_2.
\end{equation}
Using \eqref{energies}, we obtain
\begin{equation}
\bar{E}^+_n=E^-_n+R_1+R_2 . \label{ev2}
\end{equation}
From the above discussion, it is clear that we can further rationally extend $\bar{V}^+(x)$ and in fact continue to repeat the process and construct an hierarchy of rational potentials with solutions involving multi-indexed EOPs. This method allows us to obtain the explicit expressions for all these potentials and the multi-indexed EOPs in each iteration easily. The connection between SUSYQM and QHJF provides useful information to complete this exercise. \\
  
\noindent
{\bf 4 The QHJ connection}\\
In the QHJF the singularity structure analysis of the QMF in the complex plane, allows us to calculate the eigenvalues and eigenfunctions for a given potential $V^-(x)$  \cite{sree_wkb}, \cite{sree_bs}. From the same analysis we can get the entire information required to construct all the superpotentials associated with $V^-(x)$. The QMF is defined as
\begin{equation}
q(x)=\frac{d}{dx}\log\psi_m(x),    \label{qmf1}
\end{equation}
where $\psi_m(x)$ is the $m^{th}$ excited state of a potential $V^-(x)$ with eigenvalue $E_m$. Comparing with \eqref{suppot}, we can straight away see that 
\begin{equation}
\lim_{m\rightarrow 0} q(x) \rightarrow - W(x).  \label{qmf_susy}
\end{equation}
For a potential the QMF again is not unique and the different QMFs, $q_i(x)$, will lead to different $W_i(x)$.  The different QMFs can be constructed by doing a singularity structure analysis of the QMF, which consists of fixed and moving singularities. The knowledge of these singularities and their residues allows us to write the QMF in a meromorphic form, in terms of its singular and analytical parts. The location of the fixed singularities can be obtained from the QHJ equation\footnote{Note that substitution of \eqref{qmf1} in \eqref{qhj} gives the Schr\"odinger equation for the given potential.}, a Riccati equation,
\begin{equation}
q^2(x)+q^{\prime}(x)+E-V^-(x)=0, \label{qhj}
\end{equation}					
and coincide with the singularities of the potential. The moving singularities are first order poles\footnote{For all ES models, it has been seen that the point at infinity is an isolated singularity, which implies the QMF has finite number of moving poles in the complex domain.} and as seen from \eqref{qmf1}, correspond to the nodes of the $m^{th}$ excited state.  The quadratic nature of the QHJ equation \eqref{qhj}, results in the residues at all the poles being dual valued. Different residue combinations give rise to different QMFs. For all the conventional ES models, the choice of residue at the $m$ moving poles turns out to be unity.  For all these models, the QMF could be cast in the form
 \begin{equation}
q_i(x)= Q_i(x) + \frac{\partial_xP_{m}(x)}{P_{m}(x)}. \label{qmero1}
\end{equation}
Here, $\frac{\partial_xP_{m}(x)}{P_{m}(x)}=\sum_{i=0}^{m}\frac{1}{x-x_i}$ is the sum of all the principle parts of the individual Laurent expansions of $q_i(x)$ around the $m$ moving poles. Similarly,  $Q_i(x)$ is the sum of all the principle parts of the individual Laurent expansions of $q_i(x)$ around each fixed pole, plus its behavior at infinity. Different combinations of residues at the fixed poles and at the isolated singularity at infinity, will lead to different $Q_i(x)$ and hence different QMFs, $q_i(x)$. From \eqref{qmf_susy} and \eqref{qmero1}, we can see that in the limit $m \rightarrow 0$,
\begin{equation}
Q_i(x)=-W_i(x),   \label{qw}
\end{equation}
which gives us all the possible superpotentials associated with the potential $V^-(x)$. Thus \eqref{qmero} becomes
\begin{equation}
q_i(x)= -W_i(x) + \frac{\partial_xP_{m}(x)}{P_{m}(x)}. \label{qmero}
\end{equation}
Substitution of which in \eqref{qhj}, gives a second order differential equation for $P_m(x)$, 
\begin{equation}
\partial_x^2P_m(x)-2W_i(x)\partial_x P_m(x)+E_mP_m(x)=0, \label{qmfPeqn}
\end{equation}
which reduces to a COP differential equation after a suitable PCT and we obtain $P_m(x)=P^{\alpha_i}_m(x)$ a COP.
Substituting \eqref{qmero} in \eqref{qmf1}, gives the expression for the wave function as
\begin{equation}
\psi^-_m(x) = \exp\left (-\int W_i(x) dx \right) P_m^{\alpha_i}(x).  \label{wffinal}  
\end{equation}
Thus different $W_i(x)$ will lead to different differential equations leading to different solutions. In order to obtain physically acceptable solutions, we need to choose appropriate values of residues at the fixed poles such that the wave function, obtained using \eqref{qmf1}, is well behaved and does not diverge at the end points. Thus one combination of residues leads to physically acceptable solutions and the other combinations lead to unphysical solutions related to the deconjugacy of the Schr\"odinger equation \cite{disconj}. 
More specifically, in the limit $m \rightarrow 0$, the above equation reduces to 
\begin{equation}
\psi^-_0(x) = \exp\left (-\int W_i(x) dx \right).   \label{gs}
\end{equation}
Among the different $W_i(x)$ available from \eqref{qw}, the $W_i(x)$ which leads to normalizable ground state from \eqref{gs} is the superpotential which keeps SUSY exact. The same $W_i(x)$ in \eqref{qmfPeqn} and \eqref{wffinal} will lead to acceptable solutions of the potential $V^-(x)$. The other combinations lead to nonnormalizable ground states and hence give us superpotentials, which break SUSY between the corresponding partners. Thus for any given potential, we can construct all the possible superpotentials from the QMFs.

Interestingly for all the cases studies, it turns out that the exponential term is also the weight function, $w(x)$ {\it i.e.,}
\begin{equation}
\exp\left (-\int W_i(x) dx \right)=\psi^-_0(x) =w(x),   \label{wtfun}
\end{equation}
{\it w.r.t.,} which the orthogonal polynomials, $P_m^{\alpha_i}(x)$, are orthonormal. Thus the residue combination leading to physically acceptable solutions also gives a well behaved weight function in the orthonormality interval. The corresponding polynomials have only real zeros in the orthonormality interval which are governed by the oscillation theorem \cite{osci}. The other combinations lead to weight functions, which do not have the right asymptotic behavior at one or both the end points and therefore do not lead to well defined spectral problems. These solutions are used for the construction of rational potentials as seen from our earlier discussions. Thus we have a neat connection between the QMFs, superpotentials and the weight functions, which can be used to overcome the weight regularity problem, especially when performing higher iterations of isospectral deformation. The QHJ analysis works for the ES rational potentials too and for more details we refer to \cite{sree_wkb}. \\

\noindent
{\bf QHJF and the isospectral deformation} \\ 
It is clear from \eqref{partners} that  substituting $W_i(x)=-W_i(x)$ in the expression for $V_i^-(x)$ gives $V_i^+(x)$.This also implies
\begin{equation}
\psi^+_0(x)=\exp \left( \int W_i(x)dx\right).
\end{equation}  
The same substitution in \eqref{qmero} gives the QMF of $V^+_i(x)$ as
\begin{equation}
q_i(x)= W_i(x) + \frac{\partial_x P_{m}(x)}{P_{m}(x)}, \label{qmfvplus}
\end{equation}
which is nothing but $\tilde{W}_i(r)$ given in \eqref{wtild}. In addition, substituting  $W_i(x) = -W_i(x)$ in \eqref{qmfPeqn} gives \eqref{Peqn}, with  $R_1=E_m$. Thus by isospectrally deforming $V_i^-(x)$ and demanding that the polynomial $P(x)$ in \eqref{phi} be an $m^{th}$ degree polynomial, we are in fact constructing the QMFs associated with the shape invariant partner $V^+(x) \equiv \tilde{V}^+(x)$.   This connection allows us to use the singularity structure analysis to obtain the complete form of the generalized superpotential $\tilde{W}_i(r)$. The details of these will be discussed in the following sections.  \\

\noindent
 {\bf 5.  Rational potentials associated with the radial oscillator}\\
 Before proceeding further, we present the results related to the  rational potentials, belonging to the radial oscillator family, having $X_m$ EOPs as solutions studied in \cite{sreeSIP}. We perform the next iteration of isospectral deformation and rationally extend these potentials  and also obtain their solutions.
 
\noindent
The radial oscillator potential is given by
\begin{equation}
  V(r) = \frac{1}{4}\omega^2 r^2 + \frac{\ell(\ell+1)}{r^2}, \,\, r \in (0,\infty).
\label{EQ4.1}
\end{equation}
The eigenfunctions and the eigenvalues are
\begin{equation}
\psi_n(r)= r^{l+1}\exp{(-\frac{1}{4}\omega r^2)}L_n^{l+1/2}(r) \,\,;\,\, E_n^-=2n\omega   \label{wfro}
\end{equation}
respectively with $n=0,1,2\dots$.. Here $L_n^{l+1/2}(r)$ are the classical associated Laguerre polynomials which are orthonormal {\it w.r.t.}, the weight function
\begin{equation}
w_{cop}(r)=r^{l+1}\exp{(-\frac{1}{4}\omega r^2)}      \label{copwtf}
\end{equation}
in the orthonormality interval $[0,\infty)$. 

\noindent
{\bf 5.1 First iteration of isospectral deformation of the radial oscillator}\\
The four superpotentials $W_i(r), \, (i=1,2,3,4)$ for the radial oscillator obtained from the solutions of the QHJ equation \cite{sreeSIP} and the corresponding partner potentials are given in the table 1.
\begin{table*}[htb]
\centering
\begin{tabular}{|c|c|c|c|c|c|}
\hline &&&&&\\
$k$  & $ W_i(r)$
    & $V^{-}_i(r)= W^2_i(r)-\partial_r W_i(r)$
    &  $V^{+}_i(r)= W^2_i(r) + \partial_r W_i(r) $&$a_0$&$a_1$\\  \hline &&&&&\\
 $1$  & $\frac{1}{2}\omega r-\frac{(\ell+1)}{r}$
    & $V(r)- \omega(\ell+3/2)$
    &$\frac{1}{4}\omega^2 r^2 +\frac{(\ell+1)(\ell+2)}{r^2}- \omega(\ell
            +{1}/{2})$&$\ell$ &$ \ell+1 $\\\hline &&&&&\\
 $2$  &$\frac{1}{2}\omega r  + \frac{\ell}{r}$
    & $V(r)+ \omega(\ell-1/2)$
    & $\frac{1}{4}\omega^2 r^2 + \frac{\ell(\ell-1)}{r^2} +
\omega(\ell+1/2)$&$\ell $ & $\ell-1$\\ \hline &&&&&\\
 $3$  & $-\frac{1}{2}\omega r -\frac{(\ell+1)}{r}$
    & $ V(r)+\omega(\ell+3/2)$
    &  $\frac{1}{4}\omega^2 r^2 + \frac{(\ell+1)(\ell+2)}{r^2}+\omega(\ell
            +{1}/{2})$&$\ell$ &$\ell+1$ \\\hline &&&&&\\
 $4$  & $-\frac{1}{2}\omega r  + \frac{\ell}{r} $
    &  $V(r)-\omega(\ell-1/2)$
    & $\frac{1}{4}\omega^2 r^2 + \frac{\ell(\ell-1)}{r^2}-\omega(\ell
            +{1}/{2})$&$ \ell$ &$ \ell-1$\\  \hline
\end{tabular}  
\caption{Superpotentials of the radial oscillator and the corresponding partners.}
\end{table*}
\noindent
Note that the solutions of all $V^-_i(r)$  are same as \eqref{wfro} apart from the normalizing constant. Using \eqref{SI} and $a_0$, $a_1$ listed in the table 1, we can see that each set of partners $V^{\pm}_i(r)$ are SIP.  In addition $\psi^+(r)= \psi^-(r, a_0 \rightarrow a_1)$ gives the eigenfunctions for each partner potential $V^+_i(r)$.

By substituting $W_i(r)$ in \eqref{suppot} we can construct the ground state wave functions for all $V^-_i(r)$. We can see that $W_1(r)$ leads to normalizable ground state for $V_1^-(r)$ and a nonnormalizable ground state for $V_1^+(r)$. The superpotentials, $W_i(r)$ with $i=2$ and $3$ lead to nonnormalizable ground states for both the partners. The last superpotential, $W_4(r)$ leads to nonnormalizable ground state of $V_4^-(r)$, but gives normalizable ground state for $V_4^+(r)$, since $W_4(r)=-W_1(r, \ell \rightarrow \ell -1 )$. For the construction of $\tilde{W}_i(r)$, we need $W_i(r)$ which lead to $P_m^{\alpha_i}(r)$ with no zeros in the interval $[0,\infty)$. Therefore, we use $W_i(r)$ with $i=1,2,3$ in \eqref{Peqn} and this takes care of the weight regularity problem. In table 2, we list out the $\tilde{W}_i(r)$ which lead to ES rational potentials with physically acceptable solutions.  For convenience, we write $\frac{1}{2}\omega r^2=y$ from here on.\\
\begin{table*}[htb!]
	\centering
		\begin{tabular}{|c|c|c|c|c|}
		\hline 	&&&&\\
			 $W_i(r)$  & $P^{\alpha_i}_m(r)$ & $\tilde{W}_i(r)$ & $\alpha_i$ & $R_1$\\
		\hline
		&&&&\\
		$W_1(r)$& $L^{\alpha_1}_m(-y)$ & $\frac{1}{2}\omega r-\frac{(\ell+1)}{r}+ \frac{\partial_r L^{\alpha_1}_m(-y)}{L^{\alpha_1}_m(-y)} $ & $-l-\frac{3}{2}$ & $2m\omega$\\
		\hline
		&&&&\\
		$W_2(r)$ &$L^{\alpha_2}_m(-y)$&$\frac{1}{2}\omega r  + \frac{\ell}{r}+\frac{\partial_rL^{\alpha_2}_m(-y)}{L^{\alpha_2}_m(-y)}$&$l-\frac{1}{2}$ &$2 m \omega$\\
		\hline
			&&&&\\
		$W_3(r)$ & $L^{\alpha_3}_m(y)$&$-\frac{1}{2}\omega r -\frac{(\ell+1)}{r}+ \frac{\partial_rL^{\alpha_3}_m(y)}{L^{\alpha_3}_m(y)}$& $-l-\frac{3}{2}$&$-2 m \omega$
		\\
\hline
\end{tabular}\\
\caption{The superpotentials obtained after the first isospectral deformation. Here, $m=1,2\dots$ and $y=\frac{1}{2}\omega r^2$}
\end{table*}
The explicit expressions for $P^{\alpha_i}_m(r)$ obtained by substituting different $W_i(r)$ in \eqref{Peqn} along with the condition on $R_1$ are also given. These in turn give different $\tilde{W}_i(r)$ from \eqref{wtild}.  In addition the explicit expressions for the three families of the rational potentials $\tilde{V}^-_i(r)$, obtained using \eqref{vtilminus}, are given in table 3, along with their solutions, constructed using \eqref{wf} and \eqref{energies}. These involve $L1$, $L2$ and $L3$ type  generalized $X_m$ exceptional Laguerre polynomials, whose expressions are given in table 4. For the details of the calculations we refer the reader to \cite{sreeSIP}.\\
\begin{table*}[htb]
	\centering
		\begin{tabular}{|c|c|c|c|}
		\hline
		&&&\\
	$i$	& $\tilde{V}^-_i(r)$ & $\tilde{\psi}^-_{n}(r)$ & $\tilde{E}_n$ \\
		\hline
				&&&\\
	$1$ & $\tilde{V}^-_1(r) = V^-_1(r)-2\partial_x^2(\ln L^{\alpha_1}_m(-y)) + R_1$ & $ \left( \frac{y^{(l+1)/2}\exp(-\frac{y}{2})}{L^{\alpha_1}_m(-y)}\right)L_{m,n}^{III,\alpha_1}(r)$& $2\omega(n+m) $\\
	&&&\\
				\hline
	$2$	& $\tilde{V}^-_2(r)= V^-_2(r)-2\partial_x^2(\ln L^{\alpha_2}_m(-y)) + R_1$ & $\left( \frac{y^{(l+1)/2}\exp(-\frac{y}{2})}{L^{\alpha_2}_m(-y)}\right)L_{m,n}^{I,\alpha_2}(r)$ & $ 2\omega(n+m) $\\
		&&&\\
\hline
  $3$ & $\tilde{V}^-_3(r)=V^-_3(r)-2\partial_x^2(\ln L^{\alpha_3}_m(y)) + R_1$ & $\left( \frac{y^{(l+1)/2}\exp(-\frac{y}{2})}{L^{\alpha_3}_m(y)}\right)L_{m,n}^{II,\alpha_3}(r)$ & $ 2\omega(n-m) $\\
	&&&\\
\hline
\end{tabular}
\caption{First generation rational potentials, their eigenfunctions and eigenvalues}
\end{table*}
\begin{table*}[htb]
	\centering
		\begin{tabular}{|c|c|}
		\hline
		 $\mathcal{L}^{j, \alpha_i}_{m,n}(r)$ & Expression\\
		\hline
			&\\
		$L_{m,n}^{I, \,\alpha_2}(r)$&  $L_m^{\alpha_2+1}(-y)
L_n^{\alpha_2}(y)-\frac{1}{\omega r}L_m^{\alpha_2}(-y){\partial_r
L_n^{\alpha_2}(y)}$ \\
		\hline
		&\\
		$L_{m,n}^{II, \,\alpha_3}(r)$&$ (\ell + \frac{1}{2}) L^{\alpha_3 +1}_m(y)
           L^{-\alpha_3}_n(y) + r L^{\alpha_3}_m(y)\partial_r L^{-\alpha_3}_n(y)
           $\\
				\hline
		&\\
		$L_{m,n}^{III, \,\alpha_1}(r)$& $y
L^{-\alpha_1 +1}_n(y)L^{\alpha_1}_m(-y)+(m+\alpha_1)L^{\alpha_1 -1}
_m(-y)L^{-\alpha_1}_n(y)$  \\
\hline
\end{tabular}
\caption{Explicit expression of the exceptional Laguerre polynomials}
\end{table*}
From table 3, it is clear that the normalizable wave functions for each $\tilde{V}_i^-(r)$, obtained using \eqref{wf} are of the form  given in \eqref{wf2},
\begin{equation}
\tilde{\psi}^-_{n}(r)=\frac{\exp \left(-\int W_1(r) dr \right)}{P_m^{\alpha_i}( r)} L^{j,\, \alpha_i}_{m,n}(r),   \label{tildepsin}
\end{equation}
where $L^{j,\,\alpha_i}_{m,n}(r)$ represent the three exceptional Laguerre EOPs, with $j=I,II$ and $III$ representing the $L1$, $L2$ and $L3$ type polynomials respectively given in the table 4. Here $\exp(-\int W_1(r) dr)$ is nothing but the normalizable ground state $\psi_0^-(x)$ of $V^-(x)$. Moreover
\begin{equation}
 \frac{\exp \left(-\int W_1(r) dr \right)}{P_m^{\alpha_i}(r)}=w_{i,eop}(r)     \label{weop}
\end{equation}
are the rational weight functions associated with each family of the $X_m$ exceptional Laguerre polynomials. Substituting $W_1(r)$ from table 1, we can see that
\begin{equation}
\exp \left(-\int W_1(r) dr \right) = w_{cop}(r)
\end{equation}
given in \eqref{copwtf}, which gives the relation between the weight functions of the COPs and $X_m$ EOPs. Since it is already ensured that $P_m^{\alpha_i}(r)$ for $i=1,2,3$ do not have zeros on the positive real line, the weight functions for all the three cases are well behaved in the domain of orthonormality. \\

\noindent
{\bf Nontrivial supersymmetry and conventional supersymmetry}\\
It is clear that the  partners $\tilde{V}_i^{\pm}(r)$ are not shape invariant and in addition, $\tilde{V}^+_i(r)$ has solutions in terms of COPs, while  $\tilde{V}^-_i(r)$ has solutions in terms of EOPs. Thus $\tilde{W}_i(r)$ not only broke the SUSY between the partners, but also led to different types of  solutions for the two partners. Thus it  leads to different kind of SUSY between the two partners. This is called as nontrivial supersymmetry \cite{chait}.  Similar situations were encountered in the past, where isospectral deformation lead to non-shape invariant partners, but in these cases only few eigenfunctions could be calculated analytically for the deformed partner \cite{khare_book}, \cite{sukatme_iso}. In the present case, for the deformed potential too we can construct a complete set of solutions in a closed form.  

In addition, we can in fact construct yet another superpotential, $\bar{\mathcal{W}}_i(r)$ using \eqref{suppot}, which leads to a SI partner of $\tilde{V}^-(r)$. This partner will have solutions in terms of the $X_m$ EOPs with shifted parameters and the SUSY between them is exact.  Using the definition \eqref{suppot} and putting $n=0$ in the wave functions given in table 3, we obtain this different set of superpotentials,  listed in table 5. \\

\begin{table*}[htb]
	\centering
		\begin{tabular}{|c|c|}
		\hline
&\\
$\bar{\mathcal{W}}_i(r)$ & Expression\\
\hline
&\\
$\bar{\mathcal{W}}_1(r)$& $\frac{1}{2}\omega r-\frac{(\ell+1)}{r}+\frac{\partial_rL^{\alpha_1}_m(-y)}{L^{\alpha_1}_m(-y)} - \frac{\partial_rL^{\alpha_1-1}_{m+1}(-y)}{L^{\alpha_1-1}_{m+1}(-y)}  $\\
\hline
&\\$\bar{\mathcal{W}}_2(r)$& $\frac{1}{2}\omega r-\frac{(\ell+1)}{r}+\frac{\partial_rL^{\alpha_2}_m(-y)}{L^{\alpha_2}_m(-y)} -\frac{\partial_r L^{\alpha_2+1}_m(-y)}{L^{\alpha_2+1}_m(-y)}$\\
\hline
&\\
$\bar{\mathcal{W}}_3(r)$& $\frac{1}{2}\omega r-\frac{(\ell+1)}{r}+\frac{\partial_rL^{\alpha_3}_m(-y)}{L^{\alpha_2}_m(-y)} -\frac{\partial_r L^{\alpha_3+1}_m(y)}{L^{\alpha_3+1}_m(y)}$\\
\hline
\end{tabular}
\caption{Explicit expression for $\bar{\mathcal{W}}(r)$}
\end{table*}
Here we have used the fact that exceptional Laguerre polynomial reduces to a Laguerre polynomial \cite{sasaki_sigma} for $n=0$. It is very clear that for each case $\bar{\mathcal{W}}_i(r)$ is very different from $\tilde{W}_i(r)$ obtained using isospectral deformation. The partner potentials $\bar{\mathcal{V}_i}^{\pm}(r)$ can be constructed using \eqref{partners} and we see that
 \begin{equation}
  \bar{\mathcal{W}}^2_i(r)-\bar{\mathcal{W}}^{\,\prime}_i(r) =\bar{\mathcal{V}_i}^-(r) \equiv \tilde{V}_i^-(r). 
  \end{equation}
  \begin{equation}
  \bar{\mathcal{V}_i}^+(r)=\bar{\mathcal{W}}^2_i(r)+\bar{\mathcal{W}}^{\,\prime}_i(r) 
  \end{equation}
is SI partner of $\tilde{V}^-_i(r)$ and not equivalent to $\tilde{V}^+(r)$. Thus these partners share conventional SUSY and  since $\bar{\mathcal{W}}_i(r)$ is derived from normalizable wavefunction, it naturally keeps SUSY exact between the partners. Both these partners have solutions in terms of the first generation or the single indexed EOPs. The expressions for $\bar{\mathcal{V}_i}^+(r)$ and their solutions can be obtained by shifting the potential parameters, $a_0$ to $a_1$ in the expressions for  $\bar{V}_i^-(r)$ and $\bar{\psi}^-(r)$.\\ 

Note that $\bar{\mathcal{W}}_i(r)$ and $\tilde{W}_i(r)$ belong to the set of superpotentials associated with $\tilde{V}_i^-(r)$. The second iteration of isospectral deformation allows us to explore the other superpotentials associated with the potential.\\

\noindent
{\bf 5.2 Second iteration of isospectral deformation and construction of rational potentials and EOPs with two indices  }\\
In this section, we do a second iteration of isospectral deformation and construct $\bar{W}_i(r)$, using the different $\tilde{W}_i(r)$  given in table $2$. Substituting each $\tilde{W}_i(r)$ in equations \eqref{wbar}, \eqref{phi2eqnor} we obtain the equations for $\phi_2(r)$ in each case. The details of the calculations are given below.\\

\noindent
{\bf 5.2.1 Construction of $\bar{V}^+_1(r)$ and $L3$ type multi-indexed exceptional Laguerre polynomials}\\
\noindent
For the first case, we use
\begin{equation}
\tilde{W}_1(r)=\frac{1}{2}\omega r-\frac{(\ell+1)}{r}+\frac{\partial_rL^{\alpha_1}_m(-y)}{L^{\alpha_1}_m(-y)}
\end{equation}
with $\alpha_1= -l-3/2$.  Substituting this in \eqref{wbar}  gives
\begin{equation}
\bar{W}_1(r)=\frac{1}{2}\omega r-\frac{(\ell+1)}{r}+\frac{\partial_rL^{\alpha_1}_m(-y)}{L^{\alpha_1}_m(-y)} + \phi_2(r)   \label{w}
\end{equation}
and the equation for $\phi_2(r)$ from \eqref{phi2eqnor} is
\begin{equation}
\phi^2_2(r)+2\left( \frac{\omega r}{2}-\frac{(\ell+1)}{r}+\frac{\partial_rL^{\alpha_1}_m(-y)}{L^{\alpha_1}_m(-y)}\right)\phi_2(r)- \partial _x\phi_2(r)-R_2=0.    \label{phi2eqn}
\end{equation}
We need to solve the above Riccati equation to obtain $\phi_2(r)$. Noting that $\bar{W}(r)$ is nothing but one of the QMF  of $\tilde{V}^-_1(r)$, we can obtain its complete form by fixing $\phi_2(r)$. For this we use the singularity structure analysis developed in the QHJF to solve the above equation. 

From \eqref{phi2eqn}, we can see that $\phi_2(r)$ has a simple pole at $r=0$ and $2m$ fixed poles corresponding to the zeros of $L^{\alpha_1}_m(-y)$. Since we want to construct exactly solvable potentials, we continue with the ansatz that for $\phi_2(r)$ the point at infinity is an isolated singular point.\footnote{ For all ES models studied using QHJF, it was assumed that the point at infinity is an isolated singularity, which implies that the QMF has finite number of moving poles. For all models, including the ES rational potentials, this ansatz turned out to be correct \cite{sree_wkb}}. This implies $\phi_2(r)$ has finite number of moving poles and let there be $N$ such moving poles. This information allows us to write $\phi_2(r)$ as a meromorphic function in terms of its singular and analytic parts given below.
\begin{equation}
\phi_2(r)=\frac{b_1}{r}+d_1 \sum_{i=1}^{2m}\frac{1}{r-a_i} + d\,^{\prime}_1\sum_{i=1}^{N}\frac{1}{r-b_j} + c_1 r + Q(r).  \label{merophi2}
\end{equation}
Here, the first term is the principle part of the Laurent expansion of $\phi_2(r)$ around the fixed pole $r=0$, with $b_1$ being the residue. The summation terms describe the sum of all the principle parts of the individual Laurent expansions around the $2m$ fixed poles and the $N$ moving poles respectively, with $d_1$ and $d^{\prime}_1$ denoting the corresponding residues. The term $c_1r$ describes the behaviour of $\phi_2(r)$ at infinity as seen from \eqref{phi2eqn} and $Q(r)$ is the analytical part of $\phi_2(r)$, which from Liouville's theorem is a constant, say $C$. Writing $\sum_{i=1}^{2m}\frac{1}{r-a_i}=\frac{\partial_r L^{\alpha_1}_m(-y)}{L^{\alpha_1}_m(-y)}$ and $\sum_{i=1}^{N}\frac{1}{r-b_j}=\frac{\partial_r \mathcal{P}_N(r)}{\mathcal{P}_N(r)} $, where $\mathcal{P}_N(r)$ is an $N^{th}$ degree polynomial, the above equation becomes
\begin{equation}
\phi_2(r)=\frac{b_1}{r}+d_1 \frac{\partial_r L^{\alpha_1}_m(-y)}{L^{\alpha_1}_m(-y)} + d\,^{\prime}_1\frac{\partial_r \mathcal{P}_N(r)}{\mathcal{P}_N(r)} + c_1 r +C.  \label{merophi2}
\end{equation}   
In order to find the residues at these poles, we expand $\phi_2(r)$ in a Laurent expansion around each pole individually and substitute it in \eqref{phi2eqn}. For example to calculate $b_1$, we do a Laurent expansion of $\phi_2(r)$ around $r=0$,
\begin{equation}
\phi_2(r)=\frac{b_1}{r} +a_0+a_1 r+\dots     \label{b1}
\end{equation}
and substitute it in \eqref{phi2eqn}. Equating the coefficients of $1/r^2$ to zero, we obtain two values for $b_1$, namely 
\begin{equation}
b_1=0 \,\,;\,\, b_1=2l+1.
\end{equation}
Similar Laurent expansions of $\phi_2(r)$ around the fixed and moving poles and following the above method gives the dual values of residues at the $2m$ fixed and $N$ moving poles as
  \begin{eqnarray}
  d_1=0\,\,; d_1=-3, \\
	d\,'_1=0\,\,; d\,'_1=-1		\label{}
\end{eqnarray}
respectively. Next in order to calculate the behaviour of $\phi_2(r)$ at infinity, we perform a change of variable $r=1/t$ in \eqref{phi2eqn}, which gives
\begin{equation}
 \phi^2_2(t)+2\left( \frac{\omega }{2t}-(\ell+1)t-t^2\frac{\partial_rL^{\alpha_1}_m(-t)}{L^{\alpha_1}_m(-t)}\right)\phi_2(t)- t^2\partial _t\phi_2(t)-R_2=0.    \label{phi2eqnt}    
\end{equation}
The residue at $t=0$ is calculated by doing a Laurent expansion of $\phi_2(t)$ around $t=0$. Once again we obtain two values for $c_1$ as
\begin{equation}
c_1=0 \,\,; c_1=-\omega.      \label{}
\end{equation}
Substituting $\phi_2(r)$ in \eqref{phi2eqn} gives the following second order differential equation for $\mathcal{P}_N(r)$
\begin{eqnarray}
\partial_r^2\mathcal{P}_N(r)-2\left( (c_1+\frac{\omega}{2})r +\frac{b_1-\ell-1}{r} +(d_1+1)\frac{\partial_r L^{\alpha_1}_m(-y)}{L^{\alpha_1}_m(-y)}+d^{\prime}_1\frac{\partial_r \mathcal{P}_N(r)}{\mathcal{P}_N(r)}\right)\partial_r\mathcal{P}_N(r)\nonumber\\+ \left(\left[2C \left(c_1+\frac{\omega}{2}\right)r +\frac{b_1-\ell-1}{r} + \frac{\partial_r L^{\alpha_1}_m(-y)}{L^{\alpha_1}_m(-y)} \right]+2\frac{b_1}{r}\left[ \frac{\omega r}{2} + \frac{\partial_r L^{\alpha_1}_m(-y)}{L^{\alpha_1}_m(-y)}\right] \right.\nonumber\\ + \left. 2c_1 r \left[\frac{\partial_r L^{\alpha_1}_m(-y)}{L^{\alpha_1}_m(-y)} -\frac{l+1}{r}\right]C^2+2b_1c_1-c_1-R_2 \right) \mathcal{P}_N(r)=0.
\end{eqnarray}
Since all the residues are dual valued, different combinations of residues will give different equations leading to different $\mathcal{P}_N(r)$. We also know that for the construction of rational potentials, we need a $\mathcal{P}_N(r)$, which does not have zeros in the orthonormality interval. Therefore we need to choose a combination of residues, which will lead to a $\mathcal{P}_N(r)$ satisfying this requirement.  In the above equation, comparing the coefficients of $r$ gives $C=0$ and  substituting $\phi_2(r)$ in \eqref{w} gives
\begin{equation}
\bar{W}_1(r)=(c_1+\frac{\omega}{2})r +\frac{b_1-\ell-1}{r} +(d_1+1)\frac{\partial_r L^{\alpha_1}_m(-y)}{L^{\alpha_1}_m(-y)}+d^{\,\prime}_1\frac{\partial_r \mathcal{P}_N(r)}{\mathcal{P}_N(r)}.   \label{vbarqmf}
\end{equation}
Since $\bar{W}_1(r)$ is the QMF of $\tilde{V}^-_1(r)$, we need to choose residue values such that \eqref{vbarqmf} will lead to nonnormalizable solutions, more specifically to an unnormalized ground state. In other words, the choice should lead to a weight function which behaves badly at the end points, to ensure that $\mathcal{P}_N(r)$ has no zeros in the orthonormality interval, at least for certain values of the parameters. Thus the choice of residues at $r=0$ and $r=\infty$ play an important role. Since the $2m$ fixed poles do not lie in the  domain of orthonormality, we can in principle choose any of the residue values. As for the residues at the $N$ moving poles, we choose the only nontrivial value available.  For the present case, we consider the combination 
\begin{equation}
 b_1=2\ell+1;\,\, d_1=0;\,\,d\,'_1=-1;\,\, c_1=0     \label{resch}
\end{equation}
and we can check that weight function  obtained using these residues tends to infinity as $r \rightarrow \infty$. Thus we obtain
\begin{equation}
\phi_2(r) = \frac{2\ell+1}{r}- \frac{\partial_r \mathcal{P}_N(r)}{P_N(r)} + C,        \label{finalphi2}
\end{equation}
and the  equation for $\mathcal{P}_N(r)$ reduces to
\begin{equation}
\partial_r^2\mathcal{P}_N(r)-2\left(\frac{\omega r^2}{2}+\ell+r\frac{\partial_r L^{\alpha_1}_m(-y)}{L^{\alpha_1}_m(-y)}  \right)\frac{1}{r}\partial_r \mathcal{P}_N(r)+ \left(2(2\ell+1)\left(\frac{\omega}{2}+\frac{1}{r} \frac{\partial_r L^{\alpha_1}_m(-y)}{L^{\alpha_1}_m(-y)}  \right)-R_2\right)\mathcal{P}_N(r) =0.      \label{P2eqn}
\end{equation}
Performing a PCT $ \frac{1}{2}\omega r^2 = z$ and dividing the resultant equation by $2 \omega$, we obtain
\begin{equation}
z\partial_z^2 \mathcal{P}_N(z)+\left(-z-(\ell-\frac{1}{2})-2z\frac{\partial_z L_m^{\alpha_1}(z)}{ L_m^{\alpha_1}(z)}   \right)\partial_z \mathcal{P}_N(z) + \left( (2\ell+1)\left( \frac{1}{2}+\frac{\partial_z L_m^{\alpha_1}(z)}{ L_m^{\alpha_1}(z)}\right) - \frac{R_2}{2\omega}\right) \mathcal{P}_N(z) = 0.    \label{ratIII}
\end{equation}
Redefining $\ell=-d-1$ and putting $m=1$, the above equation reduces to
\begin{equation}
z\partial_z^2 \mathcal{P}_N(z)+\left(-z+(d+\frac{3}{2})-2z\frac{\partial_z L_1^{d-1/2}(z)}{ L_1^{d-1/2}(z)}   \right)\partial_z \mathcal{P}_N(z) - \left( (2d+1)\left( \frac{1}{2}+\frac{\partial_z L_1^{d-1/2}(z)}{ L_1^{d-1/2}(z)}\right) + \frac{R_2}{2\omega}\right) \mathcal{P}_N(z) = 0.   \label{P1L1}
\end{equation}
Comparing the above equation with the $X_1$ exceptional Laguerre equation of $L1$ type  given below \cite{sasaki_sigma}
\begin{eqnarray}
z\partial_z^2L^{I,\,g-1/2}_{1,n'}(z) &+& \left(-z+(g+\frac{3}{2})- 2z\frac{\partial_z L_1^{g-1/2}(z)}{ L_1^{g-1/2}(z)}   \right)\partial_zL^{I,\,g-1/2}_{1,n'}(z)  \nonumber \\  &+&\left(   -2z\frac{\partial_z L_1^{g+1/2}(z)}{ L_1^{g-1/2}(z)} + n^{\prime}+1\right)L^{I,\,g-1/2}_{1,n'}(z) =0,      \label{L1eop}
\end{eqnarray}
we can see that both the equations match, except for the last terms. Since $R_2$ is an unknown constant, we fix $R_2$ by demanding that the last terms of equation \eqref{P1L1} and \eqref{L1eop} match. This gives
\begin{equation}
R_2 = -(-n^{\prime}+d+3/2)2\omega.
\end{equation}
Though for general $m$, \eqref{ratIII} can be reduced to an $X_m$ exceptional polynomial differential equation by suitably fixing $R_2$, we find that only for $m=1$, $R_2$ turns out to be a constant. For any other choice of $m$, $R_2$ turns out to be a function of $r$, which violates the initial condition \eqref{vbar}. Thus the polynomial, $\mathcal{P}_N(r)$, coincides with 
\begin{equation}
L^{I,\,\delta}_{1,n'}(y)= L_1^{\delta+1}(-y)
L_{n^{\prime}}^{\delta}(y)-\frac{1}{\omega r}L_1^{\delta}(-y){\partial_r
L_{n^{\prime}}^{\delta}(y)},
\end{equation}
where $\delta=d-1/2$ and the degree of the polynomial is $N=n^{\prime}+1$. For small values of $n^{\prime}$ and $\ell$, we have checked numerically that as long as $-2<R_2<0$, the polynomial does not have any zeros in the interval $[0,\infty)$. A random check for bigger values of $n^{\prime}$ and $\ell$, showed the same pattern. Therefore we conjecture that as long as we choose $n^{\prime}$ and $\ell$ values such that the above condition on $R_2$ is satisfied, we have polynomials $L^{I,\,\delta}_{1,n'}(y)$ with no zeros in the orthonormality interval. Thus for all such values, we can write
\begin{equation}
  \phi_2(r)= \frac{2l+1}{r} -\frac{\partial_rL^{I,\,\delta}_{1,n'}(y)}{L^{I,\,\delta}_{1,n'}(y)},   \label{}
\end{equation}
which in turn gives 
\begin{equation}
  \bar{W}_1(r)= \frac{\omega r}{2}-\frac{\ell}{r}+\frac{\partial_r L^{\alpha_1}_1(-y)}{L^{\alpha_1}_1(-y)}- \frac{\partial_rL^{I,\,\delta}_{1,n'}(y)}{L^{I,\,\delta}_{1,n'}(y)}.     \label{}
\end{equation}
Thus $\bar{W}_1(r)$ has rational terms involving the logarithmic derivative of Laguerre polynomials and $X_1$ exceptional Laguerre polynomials of $L1$ type, belonging to the first generation EOPs. The potential $\bar{V}^+_1(r)$ is constructed by substituting  $\phi_2(r)$  in \eqref{vplusbar} and we obtain
\begin{equation}
\bar{V}^+_1(r)=V_1^+(r) +2\partial_r\left(\frac{2l+1}{r}-\frac{\partial_rL^{I,\,\delta}_{1,n'}(\frac{1}{2}\omega r^2)}{L^{I,\,\delta}_{1,n'}(\frac{1}{2}\omega r^2)}\right) + 2\omega (\ell-n^{\prime}-\frac{1}{2}),        \label{vbarcomp}
\end{equation}
which is the second generation rational extension of the radial oscillator $V^+(r)$. For different values of $n^{\prime}$ we get a different $\bar{V}^+_1(r)$. For each case, the eigenvalues can be calculated using \eqref{ev2} as
\begin{equation}
\bar{E}_n= 2\omega (n-n^{\prime}+\ell+\frac{1}{2}).
\end{equation}
Thus we can construct a family of generalized rational potentials indexed by $n^{\prime}$. Next, we calculate the solutions of $\bar{V}^+_1(r)$ using
\begin{equation}
   \bar{\psi}_{n}^+(r)=\bar{A}\bar{\psi}^-_{n}(r)\,\,;\,\, n=1,2\dots,    \label{psibarplus}
\end{equation}
where the intertwinning operators are defined in \eqref{abar}. We know $\bar{\psi}^-_{n}(r)$, since by definition $\bar{V}^-_1(r) \equiv \tilde{V}^-_1(r)$ and therefore their solutions are equal, apart from the normalization constant. Thus from tables $3$ and $4$, we get 
\begin{equation}
 \bar{\psi}^-_n(r)= \left( \frac{y^{(\ell+1)/2}\exp(-\frac{y}{2})}{L^{\alpha_1}_1(-y)}\right) L^{III,\alpha_1}_{1,n}(y)  \label{}
\end{equation}
Substituting this in \eqref{psibarplus} and operating $\bar{A}$ on it gives 
\begin{equation}
\bar{\psi}^+_n(r)=\left( \frac{y^{\ell/2}\exp(-\frac{y}{2})}{L^{\alpha_1}_1(-y)L^{I,\delta}_{1,n^{\prime}}(y)}\right)  \mathcal{Q}_{m=1, n^{\prime},n}^{\alpha_1,\delta}(y),   \label{wfbar}
\end{equation}
where
\begin{equation}
\mathcal{Q}^{\alpha_1,\delta}_{m=1, n^{\prime},n}(y)= (2\ell+1)L^{I,\delta}_{1,n^{\prime}}(y) L^{III,\alpha_1}_{1,n}(y) + r L^{I,\delta}_{1,n^{\prime}}(y) \partial_r L^{III,\alpha_1}_{1,n}(y)-r L^{III,\alpha_1}_{1,n}(y)\partial_r L^{I,\delta}_{1,n^{\prime}}(y).    \label{p1}
\end{equation}
Here, $\mathcal{Q}^{\alpha_1,\delta}_{ m=1,n^{\prime},n}(y)$ is an EOP with two indices $m=1$ and  $n^{\prime}$ taking values $1,2,\dots$, provided the condition on $R_2$ is taken care off. It can be seen that these have a complicated yet interesting structure of zeros, consisting of both the exceptional and the regular zeros. A careful investigation is needed to obtain more information regarding their distribution.   As discussed in section $4$, the nonpolynomial part in \eqref{wfbar} gives the rational weight function
\begin{equation}
w(r) = \frac{y^{\ell/2}\exp(-\frac{y}{2})}{L^{\alpha_i}_1(-y)L^{I,\delta}_{1,n^{\prime}}(y)},
\end{equation}
{\it w.r.t} which $\mathcal{Q}^{\alpha_1,\delta}_{m=1, n^{\prime},n}(y)$ are orthonormal, in the interval $[0,\infty)$. As discussed earlier, for appropriate choices of $\ell$ and $n^{\prime}$ values, these polynomials appearing in the denominator will not have any real zeros in this interval. This takes care of the weight regularity problem.  Thus we are led to a family of EOPs with two indices, which are explicitly written in terms of the single indexed EOPs. One can continue to iterate this process and construct a hierarchy of rational potentials with multi-index EOPs in their solutions. \\

\noindent
{\bf 5.2.2 Construction of $\bar{V}^+_2(r)$ and $L1$ type two indexed EOPs}\\
The second family of rational potentials is constructed by substituting 
\begin{equation}
\tilde{W}_2(r)= \frac{\omega r}{2}+\frac{\ell}{r}+\frac{\partial_rL^{\alpha_2}_m(-y)}{L^{\alpha_2}_m(-y)}
\end{equation}
 in equations \eqref{wbar} and \eqref{phi2eqnor}. The equation for $\phi_2(r)$ turns out to be
\begin{equation}
 \phi^2_2(r)+2\left( \frac{\omega r}{2}+\frac{\ell}{r}+\frac{\partial_rL^{\alpha_2}_m(-y)}{L^{\alpha_2}_m(-y)}\right)\phi_2(r)- \partial _x\phi_2(r)-R_2=0,    \label{phi2eqn2}
\end{equation}
with $\alpha_2=\ell-1/2$. As in the above case the meromorphic form of $\phi_2(r)$ is
\begin{equation}
\phi_2(r)=\frac{b_1}{r}+d_1 \frac{\partial_rL^{\alpha_2}_m(-y)}{L^{\alpha_2}_m(-y)}  + d\,^{\prime}_1\frac{\partial_r \mathcal{P}_N(r)}{\mathcal{P}_N(r)} + c_1 r + Q(r), 
\end{equation}
with $b_1$, $d_1$ and $d^{\,\prime}_1$ being residues at $r=0$, $2m$ fixed poles due to  of $L^{\alpha_2}_m(-y)$ and $N$ moving poles due to $\mathcal{P}_N(y)$. Since the point at infinity is assumed to be an isolated singularity, the analytical part $Q(r)$ is a constant $C$. Proceeding as in the previous case,  the residue values turn out to be
\begin{equation}
b_1 = 0,\,\, -2\ell-1;\, d_1=0,\,\, -3;\,d^{\prime}_1=0,\,\, -1;\,c_1 = 0,\,\,-\omega.
\end{equation}
We choose the combination $b_1=0,d_1=0,d^{\prime}_1=-1,c_1=-\omega$, such that the $\bar{W}_2(r)$ constructed leads to a nonnormalizable ground state.  With these values, we get
\begin{equation}
 \phi_2(r)=-\omega r -\frac{\partial_r \mathcal{P}_N(r)}{\mathcal{P}_N(r)} + C,    \label{}
\end{equation}
which when substituted in \eqref{phi2eqn2} gives  $C=0$ and
\begin{equation}
\partial_r^2\mathcal{P}_N(r)+2\left(\frac{\omega r^2}{2}-l-r\frac{\partial_r L^{\alpha_2}_m(-y)}{L^{\alpha_2}_m(-y)}  \right)\frac{1}{r}\partial_r \mathcal{P}_N(r)-\left(2\omega(l-\frac{1}{2}) +2 \omega r \frac{\partial_r L^{\alpha_2}_m(-y)}{L^{\alpha_2}_m(-y)}+R_2  \right) \mathcal{P}_N(r) =0.      \label{P2eqn2}
\end{equation}
Doing a point canonical transformation $\frac{1}{2}\omega r^2 =-z$  and dividing the resultant equation by $2\omega$ reduces the above equation to 
\begin{equation}
   z\partial^2_z \mathcal{P}_N(z) + \left(-z-(l-\frac{1}{2})-2z\frac{ \partial_z L^{\alpha_2}_m(z)}{L^{\alpha_2}_m(z)} \right)\partial_z\mathcal{P}_N(z) +  \left((l-\frac{1}{2}) +2z \frac{\partial_z L^{\alpha_2}_m(-z)}{L^{\alpha_2}_m(-z)}+\frac{R_2}{2\omega}\right)\mathcal{P}_N(z)=0.       \label{}
\end{equation}
 Redefining the potential parameter $\ell=-a-1$ and considering the special case $m=1$, the above equation reduces to the differential equation of the $L2$ type, $X_1$ exceptional Laguerre polynomial,
\begin{equation}
 z\partial^2_z \mathcal{P}_N(z) + \left(-z+(a+\frac{3}{2})-2z\frac{ \partial_z L^{\beta}_1(z)}{L^{\beta}_1(z)} \right)\partial_z\mathcal{P}_N(z) +\left( 1+n^{\prime}-2 (a+\frac{1}{2} )\frac{\partial_zL^{\beta+1}_1(z)}{L^{\beta+1}_1(z)}   \right) \mathcal{P}_N(z)=0,      \label{}
\end{equation}
where $\beta= -a-3/2$. As in the previous case, only for $m=1$ we obtain 
\begin{equation}
R_2=(a+\frac{1}{2}+n^{\prime})2 \omega,   \label{R22}
\end{equation}
a constant. Therefore,
\begin{equation}
\mathcal{P}_N(y) \equiv L^{II,\beta}_{1,n^{\prime}}(-y)=  (a+\frac{1}{2}) L^{\beta +1}_1(-y)
           L^{-\beta}_{n^{\prime}}(-y) + r L^{\beta}_1(-y)\partial_r L^{-\beta}_{n^{\prime}}(-y)  
\end{equation}
 with the degree $N=1+n^{\prime}$. Here again, we have checked numerically that for small values of $n^{\prime}$ and $\ell$, these polynomials will not have any zeros in the orthonormality interval, provided the potential parameters satisfy certain conditions. We have observed that the above polynomial has no zeros in the orthonormality interval,
\begin{enumerate}
\item if $R_2 \geq -1.5$ for both $n^{\prime}$ and $\ell$  being odd and 
\item if $R_2 \leq -2.5$ for both  $n^{\prime}$ and $\ell$ being even.
\end{enumerate}
Therefore by taking suitable parameter values we can obtain $\mathcal{P}_N(z)$ with the required behaviour. Thus we get
\begin{equation}
\phi_2(r)= -\omega r - \frac{\partial_r L^{II,\beta}_{1,n^{\prime}}(-y)}{L^{II,\beta}_{1,n^{\prime}}(-y)} 
\end{equation}
and
\begin{equation}
 \bar{W}_2(r)=-\frac{\omega r}{2}+\frac{\ell}{r} +\frac{\partial_r L^{\alpha_2}_1(-y)}{L^{\alpha_2}_1(-y)}-\frac{\partial_r L^{II,\beta}_{1,n^{\prime}}(-y)}{L^{II,\beta}_{1,n^{\prime}}(-y)}.     \label{}
\end{equation}
Using $\bar{W}_2(r)$, the new rational potential can be constructed from \eqref{vplusbar} as
\begin{equation}
\bar{V}^+_2(r)=  V^+(r) - 2 \partial_r \left(  \frac{\partial_r L^{II,\beta}_{1,n^{\prime}}(-y)}{L^{II,\beta}_{1,n^{\prime}}(-y)}\right) +2 \omega(n^{\prime}+a+\frac{1}{2}),  \label{v2bar} 
\end{equation}
clearly a different rational extension of the radial oscillator. The eigenvalues calculated using \eqref{ev2} will be
\begin{equation}
\bar{E}_n= (n+n^{\prime}-\ell+\frac{1}{2})2 \omega.
\end{equation}
Thus we have another family of generalized rational potentials, indexed by $n^{\prime}$ and belonging to the second generation. The eigenfunctions $\bar{\psi}^+_n(r)$ are obtained by acting the corresponding $\bar{A}$ on $ \bar{\psi}^-_n(r)=\tilde{\psi}^-_n(r)$ obtained from  tables 2 and 3. The eigenfunctions turn out to be 
\begin{equation}
\bar{\psi}^+_n(r)=\left( \frac{y^{\ell/2}\exp(-\frac{y}{2})}{L^{\alpha_2}_1(-y)L^{II,\beta}_{1,n^{\prime}}(-y)}\right)  \mathcal{Q}_{m=1,n^{\prime},n}^{\alpha_2, \beta}(y),   \label{}
\end{equation}
where
\begin{equation}
 \mathcal{Q}_{m=1,n^{\prime},n}^{\alpha_2, \beta}(y)=\left( (2\ell+1-2y)L_{1,n}^{I,\alpha_2}(y)L_{1,n^{\prime}}^{II,\beta}(-y)- r L_{1,n^{\prime}}^{II,\beta}(-y)\partial_r L_{1,n}^{I,\alpha_2}(y) -r L_{1,n}^{I,\alpha_2}(y)\partial_r L_{1,n^{\prime}}^{II,\beta}(-y)   \right).     \label{}
\end{equation}
The above polynomials are orthonormal {\it w.r.t} to the rational weight function
\begin{equation}
w(r)=\left( \frac{y^{\ell/2}\exp(-\frac{y}{2})}{L^{\alpha_2}_1(-y)L^{II,\beta}_{1,n^{\prime}}(-y)}\right).
\end{equation}
These polynomials again have both exceptional and regular zeros and well behaved rational weight functions as long as the conditions on the parameters are taken care off. Thus we obtain another set of generalized rational potentials and  $L1$ type two indexed exceptional Laguerre polynomials.\\

\noindent
{\bf 5.2.3 Construction of $\bar{V}^+_3(r)$} and $L2$ type two indexed EOPs\\
The III family of rational potentials is constructed deforming $\tilde{V}^-_3(r)$ using 
\begin{equation}
\bar{W}_3(r)= - \frac{\omega r}{2}-\frac{\ell+1}{r}+\frac{\partial_rL^{\alpha_3}_m(y)}{L^{\alpha_3}_m(y)}+\phi_2(r)
\end{equation}
in table $2$. Following the same procedure as in the previous two cases, the equation of $\phi_2(r)$ turns out to be
\begin{equation}
 \phi^2_2(r)+2\left(- \frac{\omega r}{2}-\frac{\ell+1}{r}+\frac{\partial_rL^{\alpha_3}_m(y)}{L^{\alpha_3}_m(y)}\right)\phi_2(r)- \partial _r\phi_2(r)-R_2=0,    \label{phi2eqn3}
\end{equation}
with $\alpha_3=-l-3/2$. Again the meromorphic form of $\phi_2(r)$ is written as
\begin{equation}
\phi_2(r)=\frac{b_1}{r}+d_1\frac{\partial_rL^{\alpha_3}_m(y)}{L^{\alpha_3}_m(y)}  + d\,^{\prime}_1\frac{\partial_r \mathcal{P}_N(r)}{\mathcal{P}_N(r)} + c_1 r + C, 
\end{equation}
with $b_1$, $d_1$ and $d^{\prime}_1$ being residues at $r=0$, $2m$ fixed poles due to  of $L^{\alpha_3}_m(y)$ with  and $N$ moving poles due to $P_N(y)$ respectively. The residue values turn out to be
\begin{equation}
b_1 = 0,\,\, 2\ell+1;\, d_1=0,\,\, 3;\,d^{\prime}_1=0,\,\, -1;c_1 = 0,\,\,\omega.
\end{equation}
The suitable choice $b_1=2l+1,\,d_1=0,\,d^{\prime}_1=-1,\,c_1=0$, gives
\begin{equation}
 \phi_2(r)= \frac{2\ell+1}{r}-\frac{\partial_r \mathcal{P}_N(r)}{\mathcal{P}_N(r)} +C,    \label{}
\end{equation}
which when substituted in \eqref{phi2eqn3} gives $C=0$  and
\begin{equation}
\partial_r^2\mathcal{P}_N(r)-2\left(-\frac{\omega r^2}{2}+\ell+r\frac{\partial_r L^{\alpha_3}_m(y)}{L^{\alpha_3}_m(y)}  \right)\frac{1}{r}\partial_r\mathcal{P}_N(r)+ \left(\frac{2(2l+1)}{r}\left(-\frac{\omega r}{2} +\frac{\partial_r L^{\alpha_3}_m(y)}{L^{\alpha_3}_m(y)}\right) -R_2 \right)   \mathcal{P}_N(r) = 0.      \label{P2eqn3}
\end{equation}
A point canonical transformation $\frac{1}{2}\omega r^2 =-z$ and redefining $ \ell=-b-1$ and taking the special case of $m=1$, reduces the above equation to the $L1$ type exceptional Laguerre differential equation 
\begin{equation}
 z\partial^2_z \mathcal{P}_N(z) + \left(-z+(b+\frac{3}{2}) - 2z\frac{ \partial_z L^{b-1/2}_1(-z)}{L^{b-1/2}_1(-z)} \right)\partial_z\mathcal{P}_N(z) + \left(2z \frac{ \partial_z L^{b+1/2}_1(-z)}{L^{b-1/2}_1(-z)}+n^{\prime} - 1 \right) \mathcal{P}_N(z)=0,      \label{}
\end{equation}     
and in the process fixes $R_2 = (n^{\prime}+b+3/2) 2\omega$. Thus the polynomial $\mathcal{P}_N(z))$ coincides with 
\begin{equation}
L^{I,\gamma}_{1,n^{\prime}}(-y) =  L^{\gamma +1}_1(y)L^{\gamma}_{n^{\prime}}(-y) + \frac{1}{\omega r} L^{\gamma}_1(y)\partial_r L^{\gamma}_{n^{\prime}}(-y),
\end{equation}
where $\gamma=b-1/2$ and the degree $N=1+n^{\prime}$. These polynomials do not have any zeros in the orthonormality interval with the following conditions placed on the parameters. For small values of $n^{\prime}$ and $\ell$, we have numerically checked that the polynomial will not have zeros for a given $n^{\prime}$,
\begin{enumerate}
\item if $\ell$ is even and $R_2$ is greater than $3/2$,
\item if  $\ell$ is odd $R_2$ is less than  $0$.
\end{enumerate}
Therefore for all such values we obtain 
 \begin{equation}
 \bar{W}_3(r)=-\frac{\omega r}{2}+\frac{\ell}{r} +\frac{\partial_r L^{\alpha_3}_1(y)}{L^{\alpha_3}_1(y)}-\frac{\partial_r L^{I,\gamma}_{1,n^{\prime}}(-y)}{L^{I,\gamma}_{1,n^{\prime}}(-y)}.     \label{}
\end{equation}
The new rational potential turns out to be
\begin{equation}
\bar{V}^+_3(r)=  V^+_3(r) - 2 \partial_r \left(\frac{\partial_r L^{I,\gamma}_{1,n^{\prime}}(-y)}{L^{I,\gamma}_{1,n^{\prime}}(-y)}\right) +2\omega(n^{\prime}-\ell+1/2 ),  \label{v3bar} 
\end{equation}
the third rational extension of the radial oscillator belonging to the second generation. The eigenvalues turn out to be 
\begin{equation}
\bar{E}_n= (n+n^{\prime}-\ell-\frac{1}{2})2\omega,
\end{equation}
calculated using \eqref{psibar}  \eqref{ev2}. The eigenfunctions calculated using \eqref{psibar}  with $\bar{A}=\frac{d}{dr} +\bar{W}_3(r)$ are 
\begin{equation}
\bar{\psi}^+_n(r)=\left( \frac{y^{\ell/2}\exp(-\frac{y}{2})}{L^{\alpha_3}_1(y)L^{I,\gamma}_{1,n^{\prime}}(-y)}\right)  \mathcal{Q}_{m=1,n^{\prime},n}^{\alpha_3,\gamma}(y),   \label{}
\end{equation}
where 
\begin{equation}
\mathcal{Q}_{m=1, n^{\prime},n}^{\alpha_3,\gamma}(y)= (2\ell+1-2y)L_{1,n^{\prime}}^{I,\gamma}(-y)L_{1,n}^{II,\alpha_3}(y)+L_{1,n^{\prime}}^{I,\gamma}(-y)r \partial_rL_{1,n}^{II,\alpha_3}(y)- L_{1,n}^{II,\alpha_3}(y)r\partial_r L_{1,n^{\prime}}^{I,\gamma}(-y).      \label{}
\end{equation}
 The well behaved rational weight function with respect to which these polynomials are orthonormal are
\begin{equation}
w(r)=\left( \frac{y^{\ell/2}\exp(-\frac{y}{2})}{L^{\alpha_3}_1(y)L^{I,\gamma}_{1,n^{\prime}}(-y)}\right). 
\end{equation}
These polynomials too have both exceptional and regular zeros. Thus we have a third family of generalized rational potential with $L2$ type two indexed exceptional Laguerre polynomials as solutions. \\

\noindent
{\bf 6. Discussion }\\

\noindent
{\bf The other combinations of residues} \\
In all the three cases  studied above, we have considered only one combination of the residues in obtaining the meromorphic form of $\phi_2(r)$.  Among the remaining  combinations, there will be one combination, which will lead to the superpotential $\bar{\mathcal{W}}_i(r)$ in each case.  The use of the remaining combinations of residues in $\phi_2(r)$ will lead to two different scenarios. 
\begin{enumerate}
 \item The second order differential equation for $\mathcal{P}_N(r)$ does not reduce to any of the known $X_m$ exceptional Laguerre equations. Therefore, we need to check, if the differential equation can be solved exactly and what is the nature of the solutions, $P_N(r)$, in each case.  
\item In some cases, we have seen that the differential equation for $\mathcal{P}_N(r)$ does reduce to an EOP differential equation.  In these cases, $R_2$ turns out to be a function of $r$, which violates our initial condition \eqref{vbar}. This implies we have a distinct $\bar{V}^-_i(r)$ too. These potentials may  or may not be ES and we cannot use the supersymmetric techniques to obtain the solutions of $\bar{V}^{\pm}(r)$.  
\end{enumerate}
Thus, there is a need for a careful investigation of these cases. We cannot rule out the possibility of some more rational potentials, which could be ES or quasi-exactly solvable. It will be interesting to study the nature of their solutions.\\

\noindent
{\bf New rational potentials and multi-indexed EOPs}\\
It is clear that we can continue to rationally extend the potentials constructed in this paper and obtain new ES potentials, for suitable potential parameters. These potentials will have solutions involving multi-indexed EOPs and we can explicitly construct the polynomials. In each iteration we can perform the singularity structure analysis of the Riccati equation. The appropriate choice of the residues allows us to construct the generalized superpotential. We have seen that the choice of the residues at fixed poles coinciding with the end points of the orthonormality interval are very crucial to ensure that the weight regularity problem does not arise.  

In case of the other fixed poles, lying off the orthonormality interval, we can choose any value in principle. But we see that only certain values ensure that the differential equation for $\mathcal{P}_N(r)$ reduces to a known orthogonal polynomial equation and that the resultant polynomials do not have any zeros in the orthonormality interval. Other choices may lead to new potentials, but further analysis is required to comment about the nature of these potential and their solutions.   
Thus, it is clear that with proper choice of residues, we can construct a hierarchy of rational potentials using isospectral deformation. The zeroth member in the hierarchy is the radial oscillator with COPs as solutions and is followed by a rational potentials with EOPs as solutions where the number of indices in the polynomials increases with each iteration. \\

\noindent
{\bf Nontrivial supersymmetry} \\The existence of these rational potentials adds an interesting layer to SUSYQM. Every iteration of isospectral deformation leads to superpotentials, which allow us to construct partners not related by shape invariance. The hierarchy starts with the conventional ES model, for example consider the radial oscillator with classical Laguerre polynomials in the solutions. The next member will be its rational extension with $X_m$ exceptional Laguerre polynomials in the solutions. This is followed by the another rational extension with double indexed exceptional Laguerre polynomials in the solutions and so on. Thus, nontrivial supersymmetry adds a lot of complexity and allows us to construct a hierarchy of rational ES potentials with each member very distinct from the previous one both in structure and the nature of solutions. In addition the process becomes analytically more complex with each iteration. 

In contrast to this, in the conventional supersymmetric hierarchy, we go from the radial oscillator to a shifted radial oscillator and so on, with all the potentials in the hierarchy having solutions in terms of the classical Laguerre polynomials, with shifted parameters. Similarly, if we start with a  rational radial oscillator with $X_m$ EOPs as solutions, using conventional SUSY we can construct a shifted rational potential whose solutions are in terms shifted $X_m$ EOPs. Thus, as we span the conventional supersymmetric hierarchy, we really do not get any new potentials and the process is analytically trivial.\\
%
\noindent
 {\bf 7. Conclusions}\\
In this study, we have explicitly constructed rational potentials and their eigenvalues and eigenfunctions. We have also shown that the later are in terms of the double indexed exceptional Laguerre polynomials. This method can be used to construct newer ES rational potentials with their solutions involving multi-indexed EOPs. The method is simple and makes use of the supersymmetric machinery and inputs from the QHJF, which involves the singularity structure analysis of the generalized superpotential. The fact that the isospectral deformation of a potential leads to the QMF of its shape invariant partner has been a crucial input  to arrive at the complete form of the new generalized superpotentials $\bar{W}_i(r)$.  We have shown that the appropriate choices of residues will automatically take care of the weight regularity problem and we obtain a complete set of well behaved solutions. This singularity structure analysis allows us to fix the rational terms, which extend the original potentials and makes our method different from the existing ones. 

The same method can be used to rationally extend other ES models having Hermite and Jacobi COPs and first order EOPs as solutions.  This study is currently underway and will be published elsewhere. In addition the existence of the multi-indexed EOPs leads to a lot of questions about the classification of the orthogonal polynomials, their impact on the Bochner's theorem, Sturm-Liouville's theory, ES quantum mechanical models and other related areas. Therefore a systematic study of these  new polynomials and potentials is required. \\

{\bf Acknowledgements} The author acknowledges financial support from the Science and Engineering Research Board (SERB) under the extramural project scheme, project number EMR/2016/005002. The author thanks A K Kapoor, S Rau and T Shreecharan for useful discussions and the researchers who make their manuscripts available on the arXiv and the people who maintain this archive.
\\

\end{document}